\documentclass[prl,aps,preprintnumbers,footnoteinbib,twocolumn]{revtex4-1}

\usepackage{amscd}
\usepackage{amsmath, bbm} 
\usepackage{amsfonts}
\usepackage{amssymb}
\usepackage{slashed}
\usepackage{multirow}
\usepackage{array}
\usepackage{xcolor}
\usepackage{hyperref}
\usepackage{axodraw2}

\newcommand{\Q}{\mathbb{Q}}

\begin{document} 
\title{Anomaly cancellation with an extra gauge boson}
\author{B C Allanach} \email{B.C.Allanach@damtp.cam.ac.uk} \affiliation{DAMTP, University of Cambridge, Wilberforce Road, Cambridge, 
CB3 0WA, United Kingdom}
\author{Ben Gripaios} \email{gripaios@hep.phy.cam.ac.uk} \affiliation{Cavendish Laboratory, University of Cambridge, J.J. Thomson
Avenue, Cambridge, CB3 0HE, United Kingdom}
\author{Joseph Tooby-Smith} \email{jss85@cam.ac.uk} \affiliation{Cavendish Laboratory, University of Cambridge, J.J. Thomson
Avenue, Cambridge, CB3 0HE, United Kingdom}
\begin{abstract}
Many extensions of the Standard Model include an extra gauge
boson, whose couplings to fermions are constrained by the requirement
that anomalies cancel.
We find a general solution to the resulting
diophantine equations in the plausible case where the chiral fermion content is
that of
the Standard Model
plus 3 right-handed neutrinos.
\end{abstract}
\maketitle

\section{Introduction}
Given the existence of a heavy, neutral gauge boson in the Standard
Model (SM) of particle physics -- the $Z$ boson -- it is natural to ask
whether there may be others. Such a $Z^\prime$, which corresponds to
adding an additional $u(1)$ to the gauge Lie algebra $su(3) \oplus su(2)
\oplus u(1)$ of the SM, has featured in many models extending the SM~\footnote{Examples include models of dark
  matter, axions, proton
  stabilization, the anomalous magnetic moment of the
  muon, anomalies in $B-$meson
  decays and of fermion
  masses. For a review of $Z^\prime$ physics, see~\cite{Langacker:2008yv}.} and has been the target
of   myriad  
experimental 
searches~\cite{pdg}. The
couplings of the $Z^\prime$ to fermionic matter (such as the quarks
and leptons of the SM, as well as right-handed neutrinos) cannot be
arbitrarily chosen; just as for the $Z$, we expect firstly that they
should be commensurate (corresponding to the expectation that
the gauge group is compact, although we note that we can solve for
the non-commensurate case which we discuss briefly in the
  closing remarks) and secondly that anomalies (which would
spoil the consistency of theory at the quantum level~\footnote{We
  consider here local anomalies, which require us to specify only the
  Lie {\em algebra}; consideration of global anomalies requires specifying
  the Lie {\em group} and so is more model dependent. For more details
  and some examples, see \cite{Davighi:2019rcd}.}) should
cancel~\footnote{ Regarding the model as an effective field theory,
it is possible that anomalies are compensated by a Wess-Zumino term. We shall
ignore this possibility.}.
The former implies that the fermion
charges under the extra $u(1)$ can be taken to be integers (any overall real factor can be
absorbed into the gauge coupling)
and the latter implies that
 they solve the homogeneous polynomial equations
\begin{subequations}\label{Fanomaly}
\begin{align}
 &0 =\sum_{i=1}^3\left(6Q_i\right.+3U_i+3D_i\left.+2L_i+E_i+ 
N_i \right),\label{g2X}\\
 & 0 =\sum_{i=1}^3\left(3Q_i+L_i\right),\label{22X}\\
 & 0 = \sum_{i=1}^3\left(2Q_i+U_i+D_i\right), \label{32X}\\
 & 0 =\sum_{i=1}^3\left(Q_i\right.+8U_i+2D_i\left.+3L_i+6E_i\right),\label{Y2X}\\
 & 0 =\sum_{i=1}^3\left(Q_i^2-2U_i^2+D_i^2-L_i^2+E_i^2\right),\label{YX2}\\
 &0 =\sum_{i=1}^3\left(6Q_i^3\right.+3U_i^3+3D_i^3 \left.+2L_i^3+E_i^3+
N_{i}^3\right).\label{X3}
\end{align}
\end{subequations}
Here we have assumed that the chiral~\footnote{A standard argument
 shows that vector-like fermions make no
  contrbution to the anomaly.} fermions (all of which, via charge conjugation, may be taken to
have the same chirality) consist of the just 3 SM
families of
quarks and leptons, together with 3 right-handed neutrinos, whose charges we
label by $Q_i,U_i,D_i,L_i,E_i,N_i$, respectively, with $i \in \{1,2,3\}$.
We
consider this to be the most plausible scenario, on the grounds of both
aesthetics and observation ({\em e.g.} the fit to neutrino oscillation
  data), and so we postpone comment on other possibilities to the 
    closing remarks.

Finding {\em any} solutions to diophantine equations (or even
  establishing their existence or otherwise) is, in general, a
  notoriously difficult problem in number theory (very roughly, the state of the art is a single cubic in 3
  unknowns). Surprisingly,  we will see that one can, in fact, find
  {\em all}
  solutions to (\ref{g2X}-\ref{X3}), using the sort of arithmetic
  and geometric constructions that one learns (or once learned!) in {\em kindergarten}.
  These solutions inform models where the rank of the SM is increased,
  since the extra $u(1)$ may be a sub-algebra of some larger additional gauge
  extension, as well as future phenomenological $Z^\prime$ studies.
\section{Sketch of the solution}
The keys to solving (\ref{g2X}-\ref{X3}) are twofold. The first
is to convert it to a problem in geometry by observing that one can
equivalently seek rational solutions (since any integer solution
trivially defines a rational solution and since, by clearing denominators, every
rational solution defines an integer solution). The rational numbers
form a field, allowing one to carry out division and hence various
basic geometrical constructions. The 18 charges appearing in
(\ref{g2X}-\ref{X3}) then form co-ordinates for the affine space
$\Q^{18}$. In fact, given that scaling all charges by a common
multiple leads to the same physics (as we have remarked, the scaling can be absorbed in a
redefinition of the gauge coupling), it is convenient to consider not
the charges themselves, but the equivalence classes under such a
scaling, which define the projective space $P\Q^{17}$ (whose
  points we sometimes call {\em rational points} for emphasis). The homogeneous
polynomials (\ref{g2X}-\ref{X3}) define a projective variety in
$P\Q^{17}$ whose points, which we call {\em rational solutions}, we
seek.

The second key to solving the problem is that it is easy enough
  to find {\em some} rational solutions, ({\em e.g.} by
  means of a numerical scan~\cite{Allanach:2018vjg}); 3 such points,
  $A$, $B$, and $C$, are defined in
  Table~\ref{tab:egs}.  These can be
  used as the starting point for geometric constructions. To give an
  example, consider just the quadratic (\ref{YX2}) and
suppose we know one rational point on the quadratic, $C$ say. Ignoring degenerate cases for now, a
line $L$ through $C$ intersects the quadratic at 1 other rational
point $R$ and moreover every rational point on the quadratic (indeed every
point in the ambient space!) lies on a line through $C$. Thus, by
parameterising all such lines, all rational points on the quadratic
may be found~\footnote{These arguments are standard ones in elementary number
theory~\cite{Mordell_1969}, but skeptical readers will hopefully be convinced by the explicit
  discussion that follows.}.
\begin{table*}
  \begin{tabular}{|c|ccc|ccc|ccc|ccc|ccc|ccc|}\hline
 &$Q_1$ &$Q_2$ &$Q_3$ & $U_1$&$U_2$ & $U_3$& $D_1$ &$D_2$ &$D_3$ & $L_1$&
    $L_2$&$L_3$  & $E_1$&$E_2$ &$E_3$ & $N_1$&$N_2$ &$N_3$ \\ \hline
    $A$ & 0 & 0 & 1 & 0 & 0& -4 & 0 & 0 & 2 & 0 & 0 & -3 & 0 & 0 & 6 & 0 & 0 &
    0 \\
    $B$ & 1 & 1 & 1 & -1 & -1 & -1 & -1 & -1 & -1 & -3 & -3 & -3 & 3 & 3 & 3 & 3
    & 3 & 3 \\
    $C$ & -1 & 0 & 1 & -1 & 0 & 1 & -1 & 0 & 1& -1 & 0 & 1 & -1 & 0 & 1 & 0 & 0 &
    0 \\
    \hline  \end{tabular}
  \caption{\label{tab:egs} Sample solutions of
    (\ref{g2X}-\ref{X3}). Point $A$ corresponds to the `Third Family
Hypercharge Model'~\cite{Allanach:2018lvl}, while $B$ is the
combination of baryon minus lepton number.
} 
\end{table*}

To solve the full set of Eqs. (\ref{g2X}-\ref{X3}) will require a
  more elaborate construction, as follows. Firstly, we note that the 4
  linear equations (\ref{g2X}-\ref{Y2X}) simply define a projective subspace of $P\Q^{17}$
  isomorphic to $P\Q^{13}$, to which we restrict our attention in what
  follows. Secondly, we exploit the fact that $B$ is a singular point (namely
  a point at which the underlying variety in real space is not a
  smooth manifold). In fact it is unique (up to the addition of a
  multiple of the hypercharge)~\footnote{ If one were to add multiples of
      hypercharge to any solution, one would obtain another solution. This
      redundancy could be removed, resulting in the projective dimension of
      the variety being one fewer.} among such points in that it
  is a double point of both the quadratic (\ref{YX2}) and
  the cubic (\ref{X3}). Particle physics {\em cognoscenti} will
  instantly recognize point $B$ as the combination of baryon number minus
  lepton number. (As we describe in \cite{future}, which studies how such
  singular points arise in gauge theories in general, this turns out
  to be no surprise.)

The utility of the point $B$ is the following. Since it is a double point of
the {\em cubic}, lines through it will have similar properties to the
lines through the (regular) point $C$ of the quadratic that we have
already discussed:
generically, a
line $M$ through $B$ will intersect the cubic in at 1 other rational
point, $X$ say, and moreover every rational point on the cubic (indeed every
point in the ambient space) will lie on a line through
$B$~\footnote{This observation goes back at least to Fermat and
  probably all the way to the diophantine school~\cite{Stillwell}.}.

Now let us consider the cubic and the quadratic in tandem. If $B$ were
merely a regular point of the quadratic, we would face the difficulty
that the point $X$ on the cubic would not normally lie on the quadratic. But
because $B$ is also a double point of the quadratic, we are guaranteed
that the line either lies entirely in the quadratic, or has no
point in the quadratic other than $B$. On its own, this fact is not
particularly useful, since it is the latter type of line which is
generic (consider, {\em e.g.}, the variety in $P\Q^2$ defined using
coordinates $(x,y,z) \in \Q^3$ by $xy=0$, which has a double point at
$(0,0,1)$). What is needed is a construction which generically spits out lines of
the former type. But this is easy: we use the original construction of rational
points $R$ of the quadratic, and then consider, for each such $R$, the
line $M$ joining $B$ to $R$. Generically, $R$ is distinct from $B$,
in which case
the line lies entirely in the quadratic (since it has a
point on the quadratic, {\em viz. $R$}, which is not $B$, every point
on it must be on the quadratic) and by finding the line's
other intersection with the cubic, we get a new rational solution. A
moment's consideration shows that all rational solutions of
(\ref{g2X}-\ref{X3}) can be obtained in this way.

\begin{figure}
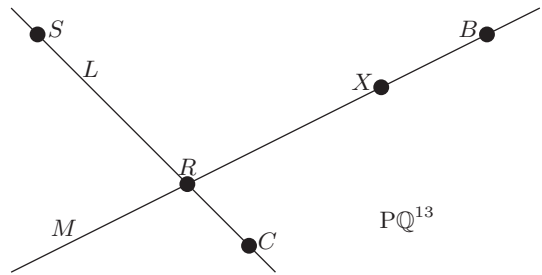

\begin{center}
  \begin{axopicture}(200,100)
    \Line(0,0)(200,100)
    \Line(0,100)(100,0)
    \Vertex(66.666,33.333){3}
    \Vertex(10,90){3}
    \Vertex(90,10){3}
    \Vertex(180,90){3}
    \Vertex(140,70){3}
    \Text(17,92){$S$}
    \Text(97,12){$C$}
    \Text(66.666,40){$R$}
    \Text(133,72){$X$}
    \Text(173,92){$B$}
    \Text(20,17){$M$}
    \Text(30,77){$L$}
    \Text(150,20){$\mathrm{P}\mathbb{Q}^{13}$}
    \end{axopicture}
\end{center}
\caption{Sketch of the geometric construction. $S$ is any
  point in the space $\mathrm{P}\mathbb{Q}^{13}$ defined by the linear anomaly
  cancellation equations, $C$ is
any point in $\mathrm{P}\mathbb{Q}^{13}$ satisfying the quadratic
equation and $B$ is the double point of both the quadratic and the
cubic equation. $L$ is the line $CS$, which generically intersects the
quadratic at $R$. $M$ is the line $BR$ which lies in the quadratic and generically intersects the
cubic at $X$, yielding a solution to all anomaly cancellation
equations. \label{fig:meth}}
\end{figure}

In summary, we have the following construction, which is shown
schematically in Fig.~\ref{fig:meth}. Starting from a rational point on the
quadratic (we take $C$, but almost any point on the quadratic distinct from $B$ would do), we
construct the line $L$ joining $C$ to an arbitrary point $S$ in
$P\Q^{13}$. This line generically hits the quadratic at a point $R$
and the line $M$ joining $R$ to the singular point $B$ (which lies in
the quadratic) generically hits the cubic at a point $X$, which is a
solution of (\ref{g2X}-\ref{X3}). Varying the position of the
point $S$ generates all
solutions, so $S \in P\Q^{13}$ parameterizes the space of solutions.

Before delving into the nitty-gritty of the parameterization, a couple of
remarks are in order. One is that we must, at some point, deal with the non-generic
cases. In the construction of solutions to the quadratic, we may find
that the line $L$ either lies entirely in the quadratic, or is tangent
to it at $C$, meaning no further solution is obtained. The same
situation may
arise for the line $M$. As we will see, they do not cause any serious
headaches. The other remark is that our parameterization of
the general solution via points $S \in P\Q^{13}$ is clearly
redundant. For example, many points $S$ will specify the same line
$L$. As we shall discuss, these redundancies could easily be removed, but
would result in uglier formul\ae.

\section{Nitty-gritty of the solution}

Given 3 points $P,P^\prime,P^{\prime \prime}$  in $P\Q^{17}$ whose
homogeneous co-ordinates are $(Q_i,U_i,D_i,L_i,N_i,E_i)$,
$(Q_i^\prime,U_i^\prime,D_i^\prime,L_i^\prime,E_i^\prime,N_i^\prime)$, and
$(Q_i^{\prime\prime},U_i^{\prime\prime},D_i^{\prime\prime},L_i^{\prime\prime},E_i^{\prime\prime},N_i^{\prime\prime})$
respectively, it will be useful to define
 \begin{align}\label{eq:q}
  q(P,P^\prime)&:=\sum_{i=1}^3\left(Q_iQ^\prime_i-2{U_i}{U_i}^\prime+{D_i}{D_i}^\prime\right. \nonumber\\
 & \left.-{L_i}{L_i}^\prime+{E_i}{E_i}^\prime\right), 
 \end{align}
 and
 \begin{align}
&c(P,P^\prime, P^{\prime \prime}):=\sum_{i=1}^3\left(6{Q_i}{Q_i}^\prime {Q_i}^{\prime\prime}+3{U_i}{U_i}^\prime {U_i}^{\prime\prime}+3{D_i}{D_i}^\prime {D_i}^{\prime\prime}\right.\nonumber \\&\left.+2{L_i}{L_i}^\prime {L_i}^{\prime\prime}+{E_i}{E_i}^\prime {E_i}^{\prime\prime}+{N_i}{N_i}^\prime {N_i}^{\prime\prime}\right).
\end{align}

Now, to find the point $R$, we
take a general point on the line $SC$, parameterized using
  homogeneous coordinates as $L = \alpha C+\beta S$, where
$\alpha,\beta \in\Q$, and substitute into (\ref{YX2}), yielding
\begin{align}
\beta(2q(C,S)\alpha+q(S,S)\beta)=0.
\end{align}
Cancelling the factor of $\beta$ (which
  appears because the point $C$ is a solution)
the general solution to this equation is
\begin{align}\label{R_expression}
R=q(S,S) C-2 q(C,S) S+\delta_{q(S,S),0}\delta_{q(C,S),0}(a C+b S),
\end{align}
where the Kronecker deltas 
(defined as $\delta_{x,y}=1$ if $x=y$ and
$\delta_{x,y}=0 \forall x \neq y$)
 encode the cases where the line lies entirely within the quadratic, with
$a,b \in \Q$ being arbitrary parameters.

To find the point $X$, we repeat the procedure, substituting the
  parameterization \ $M=\epsilon R + \gamma B$, where $\epsilon,\gamma
  \in \Q$, into the cubic
(\ref{X3}), yielding 
\begin{align}
\epsilon^2(3c(B,R,R)\gamma+c(R,R,R)\epsilon)=0.
\end{align}
Cancelling the factor of $\epsilon^2$ (which reflects the fact that
$B$ is a double point of the cubic) yields
\begin{multline}\label{eq:c}
X=c(R,R,R)B-3c(B,R,R) R\\+\delta_{c(B,R,R),0}\delta_{c(R,R,R),0}(r B+tR),
\end{multline}
with $r,t\in \mathbb{Q}$ being arbitrary parameters.

Denoting by $S_{Q_i}$ the value of $Q_i$, \emph{etc.}, at the point $S$; the restriction of $S$ to the sub-space $P\mathbb{Q}^{13}$ defined by the linear equations (\ref{g2X}-\ref{Y2X}) can be achieved by fixing $S_{Q_3}$, $S_{U_3}$, $S_{L_3}$ and $S_{E_3}$ by the relations
\begin{align}\label{eq:Scond}
S_{Q_3}&=\frac{1}{2}\left[-2S_{Q_1}-2S_{Q_2}+\sum_{i=1}^3 (S_{D_i}+ S_{N_i})\right],\nonumber\\
S_{U_3}&=-\left[S_{U_1}+S_{U_2}+\sum_{i=1}^3 (2S_{D_i}+ S_{N_i})\right],\nonumber\\
S_{L_3}&=-\frac{1}{2}\left[2S_{L_1}+2 S_{L_2}+3\sum_{i=1}^3(S_{D_i}+ S_{N_i})\right],\nonumber\\
S_{E_3}&=-S_{E_1}-S_{E_2}+\sum_{i=1}^3 (3S_{D_i}+2 S_{N_i}).
\end{align}
Our solution is then given in terms of the $18$ parameters~\footnote{
A comment on the parameter count is in order. Generically, since
  we start with 18 affine parameters and have 6 equations, we might expect
  the solution to have only 12 affine parameters. The 4 parameters
  $a,b,r,t$ appear only in degenerate cases. Furthermore, one can show
  that it suffices to restrict to points $S$ corresponding to vectors
  orthogonal to both $B$ and $C$, which brings us down to the expected
  number. We refrain from doing so, since it complicates the
  (already baroque) formul\ae .}
\begin{multline}\label{parameters}
  S_{Q_1},S_{Q_2},S_{U_1},S_{U_2},S_{D_1},S_{D_2},S_{D_3},S_{L_1},S_{L_2},S_{E_1},S_{E_2},\\
S_{N_1},S_{N_2},S_{N_3},a,b,r,t \in \Q,
\end{multline}
where the algebraic parameterization of the solution is as in (\ref{eq:c}) and $R$ is defined in (\ref{R_expression}). All
that 
remains to write the parameterization explicitly is to substitute the charges
of $B$ and $C$ from Table~\ref{tab:egs}.
The rational solution $X$ 
is then given by
\begin{align}
Q_1&=\Gamma-\Sigma+\Lambda S_{Q_1},\nonumber\\
Q_2&=\Gamma+\Lambda S_{Q_2},\nonumber\\
Q_3&=\Gamma+\Sigma+\Lambda S_{Q_3},\nonumber\\
U_1&=-\Gamma-\Sigma+\Lambda S_{U_1},\nonumber\\
U_2&=-\Gamma+\Lambda S_{U_2},\nonumber\\
U_3&=-\Gamma+\Sigma +\Lambda S_{U_3},\nonumber \\
D_1&=-\Gamma-\Sigma+\Lambda S_{D_1},\nonumber\\
D_2&=-\Gamma+\Lambda S_{D_2},\nonumber\\
D_3&=-\Gamma+\Sigma +\Lambda S_{D_3},\nonumber\\
L_1&=-3 \Gamma-\Sigma+\Lambda S_{L_1},\nonumber\\
L_2&=-3\Gamma+\Lambda S_{L_2},\nonumber\\
L_3&=-3 \Gamma+\Sigma+\Lambda S_{L_3},\nonumber\\
E_1&=3 \Gamma-\Sigma+\Lambda S_{E_1},\nonumber\\
E_2&=3\Gamma+\Lambda S_{E_2},\nonumber\\
E_3&=3 \Gamma+\Sigma+\Lambda S_{E_3},\nonumber\\
N_1&=3 \Gamma+\Lambda S_{N_1},\nonumber\\
N_2&=3 \Gamma+\Lambda S_{N_2},\nonumber\\
N_3&=3 \Gamma+\Lambda S_{N_3}, \label{n3}
\end{align}
where  
\begin{align}
\Gamma&=c(R,R,R)+r \delta_{c(B,R,R),0}\delta_{c(R,R,R),0},\nonumber\\
\Sigma&=(-3 c(B,R,R)+t  \delta_{c(B,R,R),0}\delta_{c(R,R,R),0})
\nonumber\\&(q(S,S)+a  \delta_{q(S,S),0}\delta_{q(C,S),0}),\nonumber\\
\Lambda&=(-3 c(B,R,R)+t  \delta_{c(B,R,R),0}\delta_{c(R,R,R),0})\nonumber\\&(-2q(C,S)+b  \delta_{q(S,S),0}\delta_{q(C,S),0}).
\end{align}
This solution is provided in the ancillary directory of the {\tt arXiv} preprint
of this paper in the form of a {\tt Mathematica} notebook.

One way to check that the above parameterization captures all
  solutions is to show that it can be inverted, in the following way. For a known solution
  $T$ an inverse is a set of the $18$ parameters (\ref{parameters})
  which return $T$ when substituted into~(\ref{n3}). One choice of
  parameters which achieves this is $S=T$ and, $a=0$, $b=1$, $r=0$ and
  $t=1$ ($a$, $b$, $r$ and $t$ are only needed when $T$
  corresponds to one of the exceptional cases). This inverse has
  been successfully checked on the $21\,549\,920$ solutions obtained
  by a scan in~\cite{Allanach:2018vjg}, which includes all integral solutions (up to permutations) with a maximum absolute 
charge up to $10$.

\section{Closing Remarks}
Our general solution (\ref{n3}) to Eqs. (\ref{g2X}-\ref{X3}) exploits the
  presence of a singular point, namely the one corresponding to baryon
  minus lepton number, which is unique (up to the addition of a multiple of the hypercharge) in that it is a double point of
  both the quadratic (\ref{YX2}) and the cubic (\ref{X3}). As such,
  one cannot expect the method to be of general applicability in
  studying anomaly cancellation in gauge theories. But it nevertheless
  generalizes to some situations that may be of phenomenological
  interest. A first generalization is to consider an arbitrary number $n$
  of right-handed neutrinos (RHN). Here, it turns out that our method can be
  applied provided that $n$ is odd and $n \neq 1$, with the charges of
  the extra
  neutrinos at the required
  singular point being given by $N_{2i} = +3, N_{2i+1} = -3$, for $i
  \geq 2$. It also generalizes to an odd number of SM
  families with an odd number of RHN equal to or
  exceeding the number of families, though this is probably of lesser
  phenomenological interest.  

Other cases require
other methods, but are not without hope. In Ref.~\cite{AGTS_2020b}, for example, a
related but different method was used (following Refs.~\cite{Costa_Dobrescu_Fox_2019,AGTS_2020a}) to find a complete
solution of the 1 SM family case (with an arbitrary number
of RHN) along with a number of existence
results for 3 families with a variety of numbers of RHN.

Our solution generalizes to {\em real}\/ charges,
  corresponding to the case where the gauge group is not compact. 
  The only change in our solution method would be changing rationals to reals everywhere, and as a consequence all parameters in (\ref{n3}) should be taken as real.
  Unlike in the {\em one-family}\/ SM with floating real
  hypercharges where anomaly cancellation enforces them to be commensurate~\cite{Weinberg_1995} here solutions exist
  with non-commensurate charges, for example let every SM field's charge be equal to its hypercharge and 
$N_1=\sqrt{3}$, $N_2=0$, $N_3=-\sqrt{3}$.

\section{Acknowldegments}

We thank other members of the Cambridge Pheno Working Group for discussions. This work has been partially supported by STFC consolidated grants ST/P000681/1 and ST/S505316/1. BG is also supported by King’s College, Cambridge.


\bibliography{references}

\end{document}